# Fluid dynamics–informed CCTA-derived geometric parameters in right coronary artery anomalies predict abnormal invasive Adenosine-FFR and Dobutamine-FFR

Ali Mokhtari, PhD[1†]; Anselm W. Stark, MD, PhD[1†]; Dominik Obrist, PhD[2]; Marius R. Bigler, MD, PhD[1]; Stefano F de Marchi, MD[1]; Lorenz Räber, MD, PhD[1]; Isaac Shiri, PhD[1,3]; Christoph Gräni, MD, PhD[1,3*]

[1]Department of Cardiology, Inselspital, Bern University Hospital, University of Bern, Bern, Switzerland
[2] ARTORG Center for Biomedical Engineering Research, University of Bern, Bern, Switzerland.
[3]Department of Digital Medicine, University of Bern, Bern, Switzerland

[†]Drs. Mokhtari and Stark contributed equally to this work as first authors.

[*]**Corresponding Author:**

Ali Mokhtari, PhD.
Postdoctoral Researcher Cardiac Imaging
Department of Cardiology
University Hospital Bern
Freiburgstrasse
CH - 3010 Bern, Switzerland
Email: ali.mokhtari@insel.ch

**Funding:** Swiss National Science Foundation grant number 320030 200871/1 to Christoph Gräni.

**Abstract**

**Background:** Traditional anatomical stenosis assessment may not capture the complex, eccentrically shaped lumen and its perimeter-dependent resistance in right anomalous aortic origin of coronary arteries (R-AAOCA), and may therefore underestimate flow limitation. R-AAOCA involves both fixed compression, similar to atherosclerotic plaques (assessable with invasive adenosine-derived fractional flow reserve, $FFR_{Adenosine}$), and additional stress-induced dynamic compression; both components can be captured with dobutamine-derived FFR ($FFR_{Dobutamine}$). We hypothesized that fluid dynamics–informed geometric parameters from noninvasive anatomical imaging outperform conventional metrics in predicting $FFR_{Adenosine}$ and $FFR_{Dobutamine}$.

**Methods:** We retrospectively analyzed coronary computed tomography angiography (CCTA) data of R-AAOCA patients who underwent invasive $FFR_{Adenosine}$ and $FFR_{Dobutamine}$ assessment. We calculated multiple CCTA-derived geometric parameters organized into two categories: (1) conventional geometric parameters, including cross-sectional area (A), perimeter (P), minor and major axis at the ostium, and intramural lumen area (ILA; i.e. the point of maximal stenosis), and effective diameter ($D_{eff} = \sqrt{(4A/\pi)}$); and conventional stenosis ratios including the area stenosis giratio (ASR) and effective diameter stenosis ratio (EDSR); (2) fluid dynamics–informed parameters comprising shape descriptors: including hydraulic diameter ($D_h = 4A/P$), elliptic ratio (ER = major/minor axis), circularity ($C = 4\pi A/P^2$), and hydraulic diameter stenosis ratio (HDSR); and resistance-based indices including resistance index ($RI = L \times P^4/A^4$), ostial angulation penalty ($OAP = A_{ostial}/A_{ref} \times [1 + \tan(\theta)]$), and comprehensive stenosis score ($CSS = RI \times ER \times [1 + \sin(\theta)]$) integrating resistance, shape-dependent losses, and take-off angle ($\theta$) effects. The reference was invasive $FFR_{Adenosine}$ and $FFR_{Dobutamine}$. Hemodynamic relevance was defined as FFR≤0.80 under either stress condition. We assessed Pearson correlation with FFR, univariable linear regression ($R^2$), logistic regression with odds ratios (OR), and ROC AUC for classification (FFR≤0.80).

**Results:** A total of 81 patients were included with a mean age of 52.3±12.0 years and a mean body mass index of 27.0±5.1kg/m². $FFR_{Adenosine}$≤0.80 was observed in 5 (6.2%) patients and $FFR_{Dobutamine}$≤0.80 in 16 (19.8%) patients. Highest discriminatory power to predict abnormal $FFR_{Adenosine}$ was observed for the resistance index (RI) (AUC=0.97), followed by the ostial angulation penalty (OAP; AUC=0.97), ostial area (AUC=0.96), the comprehensive stenosis score (CSS; AUC=0.95), and ostial minor diameter (AUC=0.95). Highest discriminatory power to predict abnormal $FFR_{Dobutamine}$ was observed for ostial minor diameter (AUC=0.85), followed by RI (AUC=0.83) and intramural lumen area (ILA) minor diameter (AUC=0.81). The resistance index (RI) explained 45% of the variance in $FFR_{Adenosine}$ and 43% of the variance in $FFR_{Dobutamine}$. At optimal thresholds, RI achieved 100% sensitivity and 95% specificity for predicting abnormal $FFR_{Adenosine}$, whereas ostial minor diameter achieved 100% sensitivity and 57% specificity for predicting abnormal $FFR_{Dobutamine}$.

**Conclusions:** In R-AAOCA, CCTA-derived fluid dynamics–informed metrics provide excellent and superior performance compared with conventional geometric parameters in predicting hemodynamic relevance of fixed compression. In contrast, for additional stress-induced dynamic compression, both approaches perform similarly but less accurately, highlighting the unmet need for more advanced approaches beyond static anatomy, including integrated noninvasive modeling and simulation of both fixed compression and stress-induced dynamic compression.

# Introduction

Right anomalous aortic origin of a coronary artery (R-AAOCA) is a congenital malformation in which the right coronary artery arises from the inappropriate aortic sinus of Valsalva[1,2]. The condition has been associated with myocardial ischemia, malignant arrhythmias, and sudden cardiac death, particularly in young athletes and military recruits [3–5], yet its clinical presentation is highly heterogeneous, ranging from asymptomatic incidental detection to exertional symptoms or fatal events without warning[1,2,6], making risk stratification challenging[2,7–10]. With the widespread use of coronary computed tomography angiography (CCTA), R-AAOCA is now increasingly detected as an incidental finding in adults[10,11]. The central clinical question is therefore not merely whether an anomaly is present, but whether it is hemodynamically significant and capable of producing ischemia under stress. This determination cannot be reliably made using conventional geometric parameters derived from atherosclerotic coronary disease, given the fundamentally different pathophysiology of R-AAOCA[12].

The potential hemodynamic relevance of R-AAOCA arises from a constellation of high-risk anatomical features, including an intramural course within the aortic wall, a slit-like ostium, an elliptical proximal lumen, acute take-off angles, and an interarterial course[13–15]. These anatomical features do not merely produce fixed compression; rather, they also predispose the vessel to stress-induced dynamic compression. In particular, the intramural segment may undergo progressive luminal compression and geometric distortion during exercise, when increases in heart rate, myocardial contractility, stroke volume, and aortic pressure amplify the mechanical forces acting on the anomalous coronary segment[14,16,17]. R-AAOCA therefore involves both fixed compression, similar to atherosclerotic plaques, assessable with adenosine-derived FFR ($FFR_{Adenosine}$, and additional stress-induced dynamic compression; both components can be captured with dobutamine-derived FFR ($FFR_{Dobutamine}$). This dynamic mechanism explains why functional assessment in R-AAOCA must reproduce exertional physiology as closely as possible. Although fractional flow reserve measured during adenosine-induced hyperemia ($FFR_{Adenosine}$) is the invasive reference standard for the hemodynamic assessment of fixed compression, similar to atherosclerotic plaques, it falls short in capturing stress-induced dynamic compression, as observed in R-AAOCA[13,14,18,19]. FFR under dobutamine stress ($FFR_{Dobutamine}$) therefore has particular pathophysiologic advantages, as it mimics physiological exercise conditions by increasing heart rate, myocardial contractility, and aortic pressure, thereby reproducing the hemodynamic milieu in which R-AAOCA-related ischemia (i.e. fixed compression and stress-induced dynamic compression) may occur and often yielding lower $FFR_{Dobutamine}$ values than $FFR_{Adenosine}$ in affected patients[9,19–21].

Different studies have extensively explored CCTA-derived geometric parameters to predict hemodynamic relevance in R-AAOCA. Contemporary guidelines already support a multimodality approach in which CCTA defines anatomy and physiologic testing is used when anatomical findings are concerning[22–24]. Because CCTA provides high-resolution, three-dimensional characterization of the ostium and intramural segment, it is a natural platform for non-invasive risk stratification[9,25,26]. In our previous work, we showed that approximately one in four adults with R-AAOCA had hemodynamic relevance during stress testing and that CCTA-derived geometric parameters were useful, particularly for excluding significance, but were less precise than hoped for identifying lesions that became functionally relevant by $FFR_{Dobutamine}$[10]. This is clinically important because invasive dobutamine-based assessment is procedurally demanding, requires expertise and hemodynamic monitoring, and is not ideal as a broadly applicable screening or follow-up tool[27,28]. A more accurate non-invasive method would therefore be of substantial clinical interest.

However, conventional geometric parameters assume near-circular coronary narrowing[29,30] and do not account for the perimeter-dependent viscous losses inherent to the slit-like, elliptical lumens of R-AAOCA[13,31,32]. Because hydraulic resistance in non-circular conduits depends on both cross-sectional area and perimeter[33], eccentric lesions produce disproportionately greater pressure loss than concentric ones of equivalent area reduction[34–38]. Fluid dynamics–informed parameters derived from CCTA may therefore provide a more physiologically meaningful characterization of R-AAOCA[39,40].

Accordingly, the present study tests whether fluid dynamics–informed CCTA-derived geometric parameters improve hemodynamic relevance prediction in R-AAOCA compared to conventional geometric parameters. We further evaluated whether these parameters better predict invasive $FFR_{Adenosine}$ and $FFR_{Dobutamine}$, thereby clarifying the relative contributions of fixed compression and stress-induced dynamic compression.

# Methods

**Study Design and Population**

Consecutive patients with newly identified R-AAOCA presenting to our specialized coronary artery anomaly clinic between 07/2020 and 09/2025 were prospectively enrolled in the NARCO study[14]. All included patients underwent invasive coronary angiography with $FFR_{Adenosine}$ and $FFR_{Dobutamine}$ measurements. Eligible patients were ≥18 years of age and R-AAOCA with both an interarterial and intramural course, high-quality CCTA suitable for geometric analysis, successful $FFR_{Adenosine}$ and $FFR_{Dobutamine}$ assessment, and written informed consent. Patients with anomalous left coronary artery origin were excluded because of the substantially different myocardial territory at risk and higher baseline mortality, which warrant separate investigation[1]. The study was approved by the local ethics committee (KEK 2020-00841), all participants provided written informed consent, the trial was registered at ClinicalTrials.gov (NCT04475289), and further methodological details have been reported previously[10,14].

**Intravascular Hemodynamic Assessment**

Coronary angiography was performed and pressure wire was equalized and advanced to a position distal to the intramural segment. $FFR_{Adenosine}$ and $FFR_{Dobutamine}$ measurements were subsequently performed using standardized pharmacological stress protocols, as previously described[9,41]. In brief, adenosine-induced hyperemia was achieved by continuous intravenous infusion of adenosine at 140 µg/kg/min for 2 minutes. FFR was calculated as the ratio of mean distal coronary pressure (beyond the intramural segment) to mean aortic pressure. For simulated exercise conditions, a dobutamine–volume–atropine challenge was performed: dobutamine was administered intravenously starting at 20 µg/kg/min and gradually increased, followed by at least 4 minutes at 40 µg/kg/min. To counteract preload reduction, 3 L of saline solution were administered, and 1 mg of atropine was given to achieve ≥85% of predicted maximal heart rate. $FFR_{Dobutamine}$ was measured in a hemodynamically stable state distal to the intramural segment. After data acquisition, the pressure wire was pulled back and measurements were corrected for potential drift; dobutamine was then discontinued and esmolol (40 mg bolus injections) was administered to restore heart rate to baseline. Hemodynamic relevance was defined as $FFR_{Adenosine}$ ≤ 0.80 and $FFR_{Dobutamine}$ ≤ 0.80; it was applied uniformly to allow direct comparison between stress modalities and because no alternative cutoff has been established for $FFR_{Dobutamine}$ in R-AAOCA.

**CCTA Acquisition and Geometric Analysis**

All participants underwent standard ECG-gated CCTA. Geometric measurements for the present analysis were obtained from resting diastolic phase reconstructions (typically mid-to-late diastole, 60-70%RR interval) to maximize image quality and minimize motion artifact, while acknowledging that lumen configuration may differ during systole and under stress.

Quantitative analysis focused on the ostial and proximal anomalous right coronary artery, including the intramural segment and a downstream reference segment (Figure 1). Cross-sectional measurements were obtained along the vessel using standardized anatomical definitions and measurement workflows. At each analyzed cross-section, lumen measurements included cross-sectional area (A, mm²), perimeter (P, mm; defined as the lumen boundary length), and major and minor axes ($D_{major}$ and $D_{minor}$, mm) derived from an ellipse fit to the lumen contour. These measurements were used to derive conventional and fluid dynamics–informed geometric parameters. Reproducibility of CCTA-derived anatomical measurements was assessed within the published framework, including intra- and interrater analyses reported in the supplementary methods[10,42].

**Derived Geometric Parameters**

Both conventional and fluid dynamics–informed geometric parameters were calculated from CCTA measurements. We defined "fluid dynamics–informed parameters" as geometric indices that incorporate principles from fluid mechanics, specifically, the dependence of hydraulic resistance on both cross-sectional area and perimeter in non-circular conduits, as opposed to conventional geometric parameters, which assume circular luminal geometry or rely solely on area- and diameter-based measurements. This distinction is grounded in the Hagen-Poiseuille framework, in which viscous resistance scales with the inverse fourth power of the hydraulic diameter, a perimeter-sensitive quantity that captures resistance effects absent from conventional area-derived metrics (see Table 1 for all formulas and definitions). Fluid dynamics–informed parameters were selected based on theoretical relevance to flow resistance in non-circular conduits and validated in computational fluid dynamics (CFD) studies[12,22,33,36,37,43,44].

On the contrary, conventional geometric parameters included stenosis length, cross-sectional areas, and diameters at the intramural lumen area (ILA, defined as the point of maximal stenosis), reference segment dimensions, take-off angle, and conventional stenosis ratios (ASR, EDSR). Fluid dynamics–informed parameters comprised shape descriptors including ER, circularity, and hydraulic diameter ($D_h$ = 4A/P), which accounts for the disproportionate flow resistance in non-circular conduits[33,36,43], as well as the hydraulic diameter stenosis ratio (HDSR). In addition, resistance-based indices were derived: the RI (RI = L×$P^4$/$A^4$), grounded in Poiseuille's law where resistance scales with L/$D_h^4$, the OAP, and the comprehensive stenosis score (CSS) integrating geometric resistance, shape-dependent losses, and angulation effects. A complete summary of all parameters with formulas is provided in Table 1. Definitions and physical rationale for the effective diameter, hydraulic diameter, RI, and composite scores are illustrated in Figure 2. Notably, stenosis ratios derived from $D_{eff}$ and $D_h$ diverge as lumen ellipticity increases, with the error of area-based ($D_{eff}$) estimates rising with ER. Detailed definitions are available in the Supplementary Methods.

**Statistical Analysis**

Statistical analyses were performed separately for each invasive reference standard (i.e. $FFR_{Adenosine}$ and $FFR_{Dobutamine}$). FFR was analyzed both as a continuous variable and as a binary outcome, with hemodynamic significance defined a priori as abnormal $FFR_{Adenosine}$ or $FFR_{Dobutamine}$ (i.e. $\leq 0.80$).

Descriptive statistics are presented as mean±standard deviation (SD) or median (interquartile range [IQR]) for continuous variables, depending on distribution assessed by the Shapiro–Wilk test, and as counts and percentages for categorical variables. Collinearity among CCTA-derived geometric parameters was evaluated using a Pearson correlation matrix.

Associations between individual geometric parameters and continuous FFR were assessed using Pearson correlation coefficients with 95% confidence intervals (95% CI) derived using Fisher's z-transformation. Univariable linear regression was performed to quantify the relationship between each geometric parameter and FFR, reporting regression coefficients (β), standard errors (SE), 95% CI, p-values, and coefficients of determination ($R^2$). For binary classification of hemodynamic significance, univariable logistic regression models were fitted for each geometric parameter after z-score standardization to enable comparable effect estimates. Model discrimination was assessed using ROC analysis, summarized by the AUC with 95% CI calculated using the DeLong method. Odds ratios (OR) with 95% CI were derived from fitted regression coefficients, and statistical significance was evaluated using two-sided Wald tests.

Pairwise comparisons of discriminatory performance between geometric parameters were performed using DeLong's test for comparing correlated ROC curves, which accounts for the correlation between AUC estimates derived from the same sample by computing a variance–covariance matrix based on the theory of generalized U-statistics. The test was applied to all pairwise combinations of parameters achieving AUC $\geq 0.75$ under each stress condition to determine whether any parameter demonstrated statistically superior diagnostic accuracy.

For parameters demonstrating discriminatory performance (AUC >0.75), optimal thresholds for identifying FFR $\leq 0.80$ were determined using Youden's J statistic, and corresponding sensitivity, specificity, positive predictive value (PPV), negative predictive value (NPV), and proportions of hemodynamically relevant cases above and below each cutoff were calculated. Given the exploratory nature of this analysis and the evaluation of multiple geometric parameters, p-values were not adjusted for multiple comparisons; emphasis was placed on effect sizes and 95%CI. All tests were two-sided, with statistical significance defined as p <0.05. Analyses were performed in Python (v3.12.7) using NumPy (v1.26.4), pandas (v2.3.3), scikit-learn (v1.5.1), and statsmodels (v0.14.2).

# Results

**Study Population and Baseline Characteristics**

A total of 81 patients with R-AAOCA comprised the final analytic cohort. The study population had a mean age of 52.3±12.0 years (range 22–81 years) with a body mass index of 27.0±5.1 kg/m². Complete geometric analysis was available for all participants. The intramural segment demonstrated substantial elliptical deformation, with a mean ostial ellipticity of 2.89±0.85 and ILA ellipticity of 2.46±0.72, confirming the non-circular geometry characteristic of R-AAOCA. The mean stenosis length was 10.41±4.20 mm, and the mean RI was 179.91±261.02. Circularity values at the ostium (0.70±0.14) and ILA (0.74±0.13) were below 1.0, indicating non-circular geometry. Additional baseline geometric characteristics, including all conventional and fluid dynamics–informed parameters, are detailed in the Supplementary Material.

**Hemodynamic Findings**

Under adenosine stress, hemodynamically significant stenosis ($FFR_{Adenosine} \leq 0.80$) was identified in 5 of 81 patients (6.2%). Under dobutamine stress, 16 of 81 patients (19.8%) demonstrated $FFR_{Dobutamine} \leq 0.80$. The approximately three-fold higher prevalence with $FFR_{Dobutamine}$ underscores the incremental value of dynamic stress testing in R-AAOCA.

**Predictive Performance Under Adenosine Stress**

Using $FFR_{Adenosine}$ as the reference, fluid dynamics–informed parameters showed high discriminatory performance (Figure 3). The RI and OAP achieved an AUC of 0.97. The CSS achieved an AUC of 0.95. Conventional geometric parameters also performed well: ostial area (AUC = 0.96), ostial minor diameter (AUC = 0.95), ILA area (AUC = 0.94).

*Correlation and Linear Regression.* The RI demonstrated the strongest negative correlation with FFR (r = −0.67, 95% CI: −0.77 to −0.53, p < 0.001), explaining 45% of FFR variance ($R^2 = 0.45$). Linear regression confirmed that higher RI values were associated with lower FFR (β = −0.001, p < 0.001). The CSS also showed a strong correlation (r = −0.53, 95% CI: −0.67 to −0.35, $R^2 = 0.27$, p < 0.001). Among conventional geometric parameters, ostial area demonstrated a positive correlation with FFR (r = 0.43, 95% CI: 0.23 to 0.59, p < 0.001), consistent with larger lumen dimensions being associated with preserved coronary flow (Table 2).

*Logistic Regression.* Univariable logistic regression confirmed the significant predictive value of geometric parameters for binary classification of hemodynamic significance (FFR ≤ 0.80). The RI demonstrated the strongest association (OR = 6.39, 95% CI: 1.60–25.59, p = 0.006), indicating that each SD increase in RI was associated with a more than six-fold increase in odds of hemodynamically significant stenosis. The CSS was also associated (OR = 2.61, 95% CI: 1.32–5.16, p = 0.009). Ostial minor diameter (OR = 0.01, 95% CI: 0.00–0.57, p = 0.025) and ILA minor diameter (OR = 0.04, 95% CI: 0.00–0.53, p = 0.014) showed inverse associations (Table 3).

Pairwise AUC comparisons using DeLong's test revealed no statistically significant differences among the top-performing parameters for $FFR_{Adenosine}$ (all p > 0.05; Supplementary Table S2).

**Predictive Performance Under Dobutamine Stress**

Using $FFR_{Dobutamine}$ as the reference, overall discriminatory performance was lower than for $FFR_{Adenosine}$ (Figure 3). Ostial minor diameter achieved the highest AUC (0.85), followed by RI (AUC = 0.83), CSS (AUC = 0.82), ILA minor diameter (AUC = 0.81), and OAP (AUC = 0.80) (Figure 4).

*Correlation and Linear Regression.* The RI showed a correlation with $FFR_{Dobutamine}$ (r = −0.66, 95% CI: −0.77 to −0.51, p < 0.001). The CSS showed similar results (r = −0.59, $R^2$ = 0.35). Among conventional geometric parameters, ILA minor diameter demonstrated the strongest positive correlation (r = 0.50, 95% CI: 0.32 to 0.65, $R^2$ = 0.25, p < 0.001), followed by ostial minor diameter (r = 0.47, 95% CI: 0.29 to 0.63, $R^2$ = 0.23, p < 0.001) (Table 2).

*Logistic Regression.* On univariable logistic regression, the RI demonstrated the strongest association (OR = 6.26, 95% CI: 2.17–18.06, p = 0.001). Ostial minor diameter (OR = 0.12, 95% CI: 0.03–0.44, p = 0.001) and ILA minor diameter (OR = 0.20, 95% CI: 0.07–0.58, p = 0.003) showed inverse associations. The CSS (OR = 3.47, 95% CI: 1.48–8.12, p = 0.002) and OAP (OR = 0.18, 95% CI: 0.05–0.57, p = 0.004) also demonstrated significant associations (Table 3). For $FFR_{Dobutamine}$, the discriminatory performance of RI decreased, with AUC declining from 0.97 ($FFR_{Adenosine}$) to 0.83 (Δ = −0.14).

Conventional geometric parameters based on minor diameter showed more consistent performance across stress modalities (ostial minor diameter: AUC 0.95 → 0.85, Δ = −0.10), while area-based conventional parameters showed greater attenuation (ostial area: AUC 0.96 → 0.76, Δ = −0.20). DeLong's test confirmed that ostial minor diameter demonstrated significantly higher AUC than ostial area for $FFR_{Dobutamine}$ (0.85 vs 0.76, z = 2.04, p = 0.041), the only statistically significant pairwise difference observed across all comparisons (Supplementary Table S3). All other pairwise comparisons were non-significant (Supplementary Table S3).

While discriminatory performance (AUC) decreased for $FFR_{Dobutamine}$, the explained variance ($R^2$) for the fluid dynamics–informed parameters remained relatively preserved: the RI explained 45% of FFR variance for $FFR_{Adenosine}$ versus 43% for $FFR_{Dobutamine}$ (Δ = −2 percentage points), while the CSS explained 27% versus 35% (Δ = +8 percentage points).

**Optimal Cutoff Values**

Using Youden's J statistic to maximize the sum of sensitivity and specificity, optimal classification thresholds were derived for the top-performing parameters under each stress condition (Table 5). For $FFR_{Adenosine}$, the fluid dynamics–informed indices RI and CSS achieved perfect sensitivity with excellent specificity: the RI at a cutoff of $\geq 406.4$ mm$^{-3}$ achieved 100% sensitivity and 95% specificity, while the CSS at $\geq 1068.3$ achieved 100% sensitivity and 92% specificity. Conventional geometric parameters also achieved perfect sensitivity: ostial area $\leq 4.1$ mm² (specificity 89%) and ostial minor diameter $\leq 1.10$ mm (specificity 92%).

For $FFR_{Dobutamine}$, perfect sensitivity (100%) was achieved only by ostial minor diameter at a cutoff of $\leq 1.60$ mm, with lower specificity (57%). The fluid dynamics–informed indices for $FFR_{Dobutamine}$ favored specificity over sensitivity: the RI at $\geq 284.6$ mm$^{-3}$ provided 94% specificity with 69% sensitivity, while the CSS at $\geq 796.8$ achieved 91% specificity with 67% sensitivity.

## Discussion

In R-AAOCA, CCTA-derived fluid dynamics–informed metrics provide excellent and superior performance compared with conventional geometric parameters in predicting hemodynamic relevance of fixed compression. In contrast, for additional stress-induced dynamic compression, both approaches perform similarly but less accurately, highlighting the need for more advanced approaches beyond static anatomy, including integrated modeling and simulation of both fixed compression and stress-induced dynamic compression. Fluid dynamics–informed parameters were particularly effective in predicting $FFR_{Adenosine}$, supporting the relevance of non-circular conduit resistance in this lesion characterized by eccentric, slit-like geometry. Performance was attenuated for $FFR_{Dobutamine}$, where the ostial minor diameter remained the strongest predictor, consistent with prior multimodality findings[10,41]. Taken together, these findings support a mechanistically coherent framework in which fixed compression is well captured by resting CCTA-derived geometric parameters as reflected by $FFR_{Adenosine}$, whereas stress-induced dynamic compression introduces additional physiology that resting anatomy can only partially represent[25,41].

 The present results are best interpreted as an extension of prior work rather than a replacement of established markers. The persistence of ostial minor diameter as the dominant predictor of $FFR_{Dobutamine}$ is a clinically important confirmatory finding: even after introducing more sophisticated fluid dynamics–informed parameters, the strongest single anatomical signal remained concentrated at the ostial/proximal intramural region. This is biologically plausible because R-AAOCA commonly exhibits ostial and proximal intramural elliptical deformation, and the narrow minor dimension is expected to have disproportionate hemodynamic importance in high-aspect-ratio stenotic geometries[10,35]. In practical terms, this consistency strengthens the case for prioritizing ostial minor diameter in CCTA assessment of R-AAOCA, while positioning fluid dynamics–informed parameters as complementary tools that improve physiological interpretation rather than replacing it.

The theoretical basis for fluid dynamics–informed parameter design is well established in fluid mechanics. In non-circular conduits, hydraulic resistance depends on both area and perimeter, and the hydraulic diameter provides a first-order geometric surrogate of viscous resistance.  Eccentric stenoses generate greater pressure gradients than concentric lesions of equivalent area, and with a mean ostial ellipticity of $2.89^{35,43}$, R-AAOCA falls within the range where these effects are substantial. The superior performance of fluid dynamics–informed parameters for $FFR_{Adenosine}$ is not merely an empirical observation; it is consistent with the expectation that circular simplifications introduce systematic error when applied to R-AAOCA lesions with marked eccentricity.

The RI integrates stenosis length with the perimeter-to-area relationship into a term mathematically equivalent to $L/D_h^4$, a Poiseuille-inspired geometric surrogate of viscous resistance in non-circular conduits. Its excellent discrimination for $FFR_{Adenosine}$ (AUC = 0.97 , matching the OAP) provides empirical support for the relevance of this fluid dynamics–informed framework in R-AAOCA and extends prior observations from atherosclerotic coronary disease, where morphologic composite indices incorporating lesion geometry and length have similarly outperformed conventional geometric parameters[39,40].

The differential behavior of the same geometric parameters for $FFR_{Adenosine}$ versus $FFR_{Dobutamine}$ highlights a central pathophysiologic feature of R-AAOCA: the distinction between fixed compression and stress-induced dynamic compression. $FFR_{Dobutamine}$ has emerged in the R-AAOCA literature precisely because it better reproduces exertional physiology and dynamic compression of the anomalous intramural segment than $FFR_{Adenosine}$ alone. In a recent study, invasive evaluation with intravascular ultrasound (IVUS) plus $FFR_{Adenosine}$ and $FFR_{Dobutamine}$ was explicitly framed around this dynamic compression problem, and the authors note both the complexity and expertise required for such testing[41]. Accordingly, the attenuation of resting CCTA-based geometric prediction for $FFR_{Dobutamine}$ likely reflects stress-induced dynamic compression that is not fully captured by a resting, predominantly diastolic acquisition [31].

A useful mechanistic distinction is between pulsatile lumen deformation and stress-induced dynamic compression. Pulsatile deformation refers to beat-to-beat cyclic changes in lumen shape and caliber driven by the cardiac cycle; this phenomenon has been documented by IVUS in intramural R-AAOCA under resting conditions, including dynamic flattening and cross-sectional changes across the cycle[45]. Stress-induced deformation, in contrast, refers to the additional shift in mean lumen geometry that occurs when transitioning from rest to physiological stress. Our data cannot directly separate these processes, but the modality-dependent performance pattern strongly suggests that both are relevant and that the latter is a major determinant of $FFR_{Dobutamine}$ -defined hemodynamic significance.

The underlying mechanism of stress-induced lumen compromise remains incompletely resolved, and our study was not designed to establish causality. These mechanisms remain hypothesized and likely interact. Nevertheless, the observed findings are consistent with several plausible, non-mutually exclusive mechanisms described in prior R-AAOCA work. First, mechanical compression due to aortic root expansion and pressure loading may narrow the intramural segment, which is structurally coupled to the aortic wall; IVUS studies and prior mechanistic imaging reports have described dynamic compression and phasic shape changes consistent with this concept. Second, fluid-dynamic effects, including local acceleration through a narrowed compliant segment with associated pressure reduction (often discussed in Bernoulli/Venturi terms), may amplify trans-lesional pressure loss and promote inward displacement of the vessel wall under stress. Experimental and computational R-AAOCA studies support the hemodynamic relevance of intramural-segment pressure-flow interactions, particularly in ischemic phenotypes[17,46]. Third, three-dimensional interactions among the anomalous coronary artery, aortic root, pulmonary artery, and commissural anatomy may constrain or accentuate deformation depending on patient-specific anatomy.

An important implication of the present work is that our parameters, although physiologically motivated, are insufficient to predict geometry directly. This limitation is not specific to one-parameter design; rather, it is an intrinsic constraint of resting CCTA. A single static acquisition cannot fully represent the time-varying geometry that emerges from the interaction of vessel wall mechanics, loading conditions, and surrounding structures during stress. Consequently, fluid dynamics–informed resting metrics may improve phenotyping of fixed anatomical resistance while still failing to identify all patients whose hemodynamic significance is driven predominantly by stress-induced dynamic compression. These findings therefore sharpen rather than replace the role of stress-based functional testing.

At the same time, the persistent superiority of the ostial minor diameter for predicting $FFR_{Dobutamine}$ suggests a clinically meaningful hypothesis: this measurement may encode information beyond instantaneous fixed narrowing. Specifically, a smaller resting ostial minor diameter may reflect a less mechanically stable proximal lumen configuration that is closer to a threshold for dynamic collapse and therefore more susceptible to stress-induced deformation. This interpretation is compatible with the slit-like/elliptical morphology emphasized in R-AAOCA imaging literature and with IVUS evidence of dynamic minor-axis deformation in intramural segments. Validation in studies integrating dynamic IVUS and patient-specific computational modeling will be required to confirm whether resting morphology can serve as a surrogate of deformation susceptibility.

The clinical implications of this work are twofold. First, the results support continued use of conventional geometric parameters, particularly ostial minor diameter, for noninvasive risk stratification in R-AAOCA. Notably, an ostial minor diameter >1.10 mm effectively excludes hemodynamic significance by $FFR_{Adenosine}$ (100% NPV), while a threshold >1.60 mm achieves the same for $FFR_{Dobutamine}$, a larger value consistent with the additional burden of stress-induced dynamic compression beyond fixed compression. Second, they justify the parallel development of fluid dynamics–informed parameters, which may improve mechanistic interpretation and enhance prediction where fixed geometric resistance is dominant. More broadly, our findings reinforce the importance of a staged approach in which resting anatomy informs initial triage, while stress-based physiologic testing remains necessary when the clinical concern is exertional ischemia driven by dynamic compression. In this regard, the present study is important not because it proposes a replacement for invasive $FFR_{Dobutamine}$ assessment, but because it defines a more rigorous and physiologically grounded anatomical framework on which future dynamic and noninvasive approaches can build.

## Limitations

Several limitations warrant consideration. First, this was a single-center study and CCTA measurements were obtained at rest in diastole, potentially missing stress-induced dynamic geometric changes. Standardized diastolic acquisition and formal reproducibility assessment partially mitigate this; stress-phase or real-time intravascular imaging would be the ideal approach. Second, the low prevalence of hemodynamic significance by $FFR_{Adenosine}$ (6.2%) yielded wide confidence intervals for some estimates; the consistent direction of associations across multiple independent parameters supports overall robustness, though a larger multicenter cohort with higher event rates would allow more precise estimates and multivariable modeling. Third, the Poiseuille-inspired fluid dynamics–informed indices do not fully account for flow separation, entrance effects, or nonlaminar behavior in the intramural segment; while parameters were grounded in established fluid dynamics theory and supported by published CFD evidence, patient-specific computational simulations would provide a more rigorous validation framework. Finally, cutoff values were derived and tested within the same cohort, introducing optimism bias; all thresholds are therefore reported as exploratory, and prospective validation in independent populations is required before clinical implementation.

## Conclusions

In R-AAOCA, CCTA-derived fluid dynamics–informed metrics outperform conventional geometric parameters for predicting hemodynamic relevance of fixed compression. In contrast, both approaches show similarly limited accuracy for stress-induced dynamic compression. These findings highlight that R-AAOCA cannot be assessed using concepts from atherosclerotic disease. Fluid dynamics–informed CCTA provides a more physiologically grounded evaluation of fixed compression in this eccentric, non-circular lesion, but static anatomy alone remains insufficient to capture stress-induced dynamic compression. This highlights the unmet need for more advanced approaches beyond static anatomy, including integrated noninvasive modeling and simulation of both fixed compression and stress-induced dynamic compression.

**Figures**

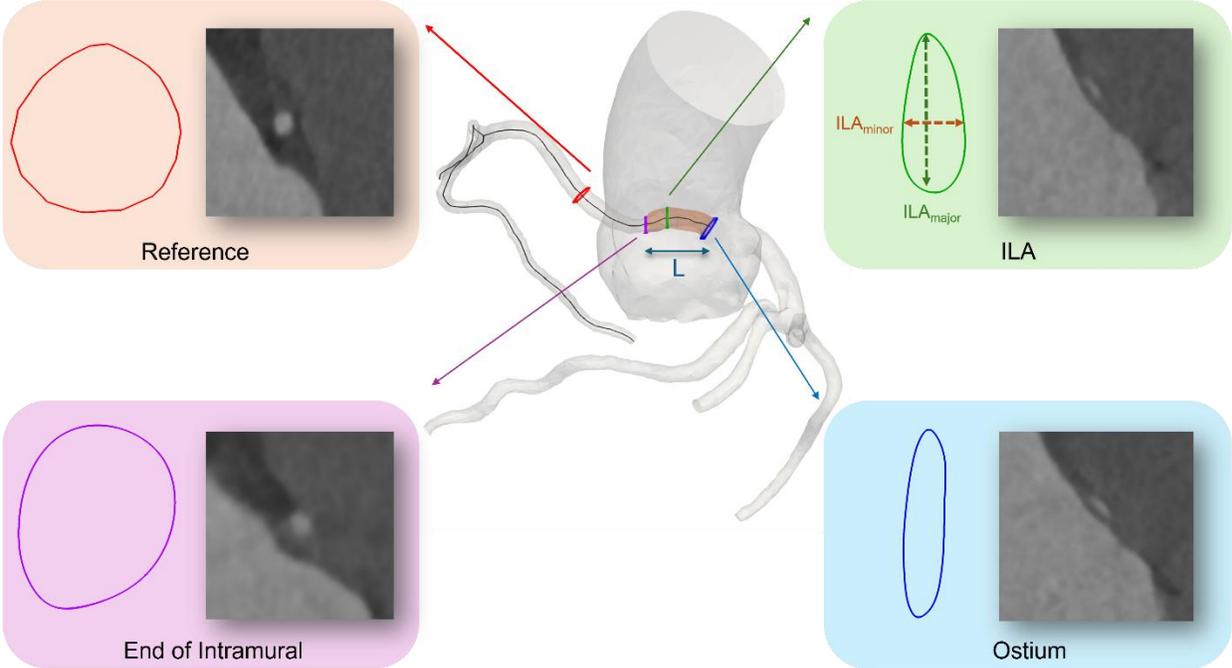

**Figure 1.** Anatomical Measurement Locations. Cross-sectional planes at ostium, intramural lumen area (ILA), end of intramural segment, and reference segment with corresponding CCTA images. L = stenosis length.

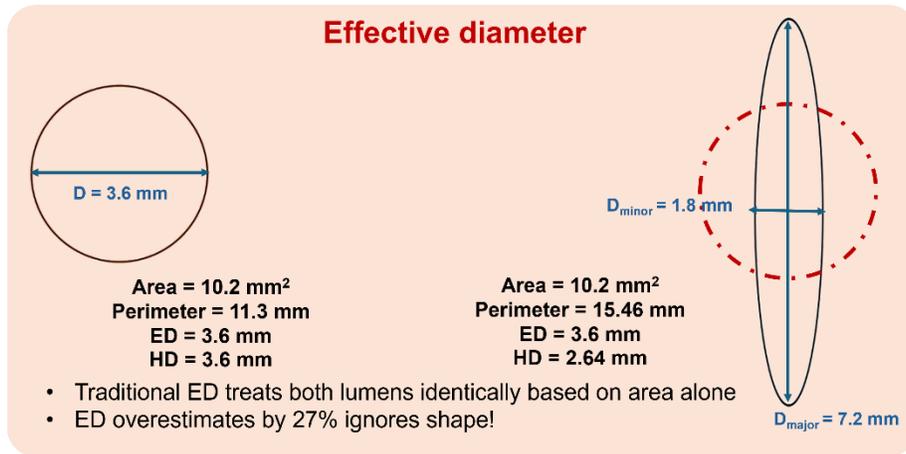
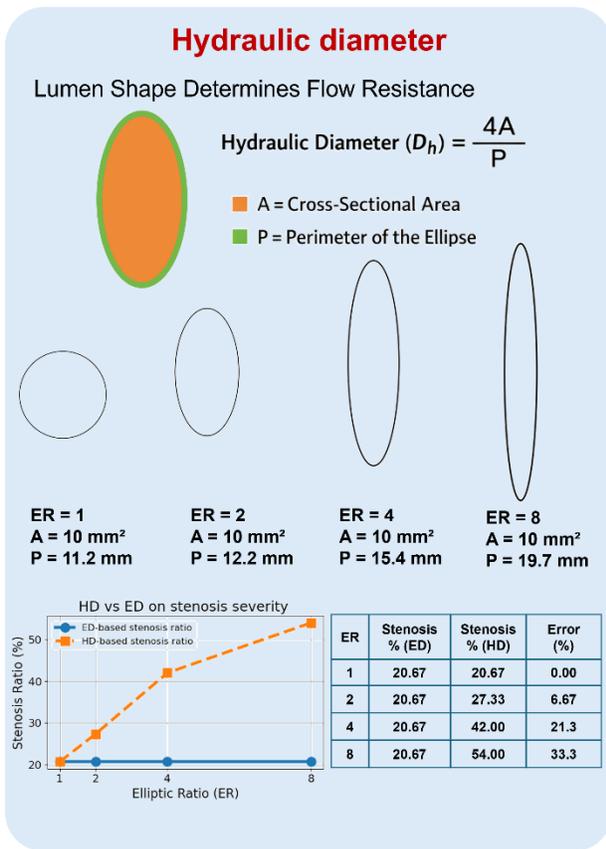
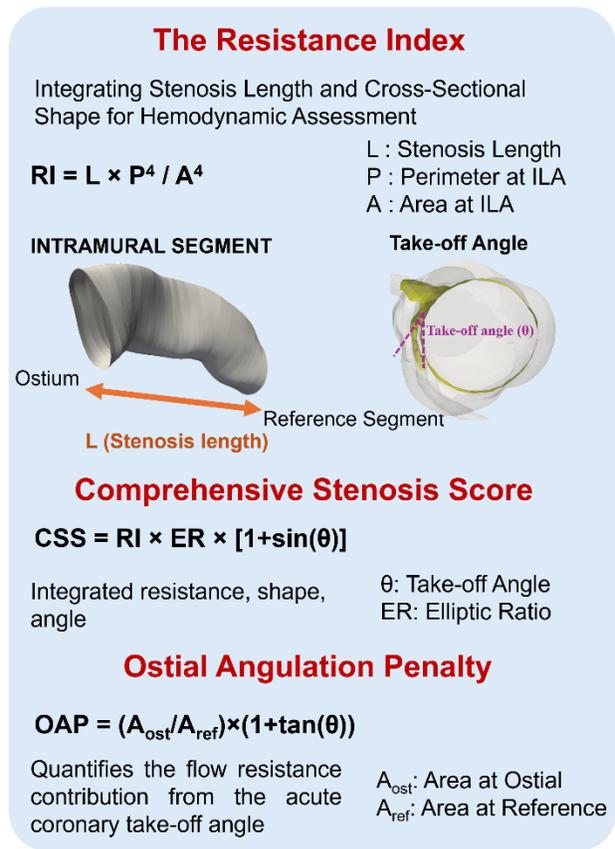

**Figure 2**. Fluid Dynamics–Informed Parameter Definitions. Hydraulic diameter ($D_h = 4A/P$) decreases with ellipticity despite constant area. Resistance index (RI), comprehensive stenosis score (CSS), and ostial angulation penalty (OAP) formulas shown. At equal area, increasing ellipticity increases perimeter, reducing $D_h$ and amplifying HD-based stenosis ratios relative to ED-based estimates. CSS, comprehensive stenosis score; $D_h$, hydraulic diameter; ED, effective diameter; HD, hydraulic diameter; OAP, ostial angulation penalty; P, perimeter; RI, resistance index.

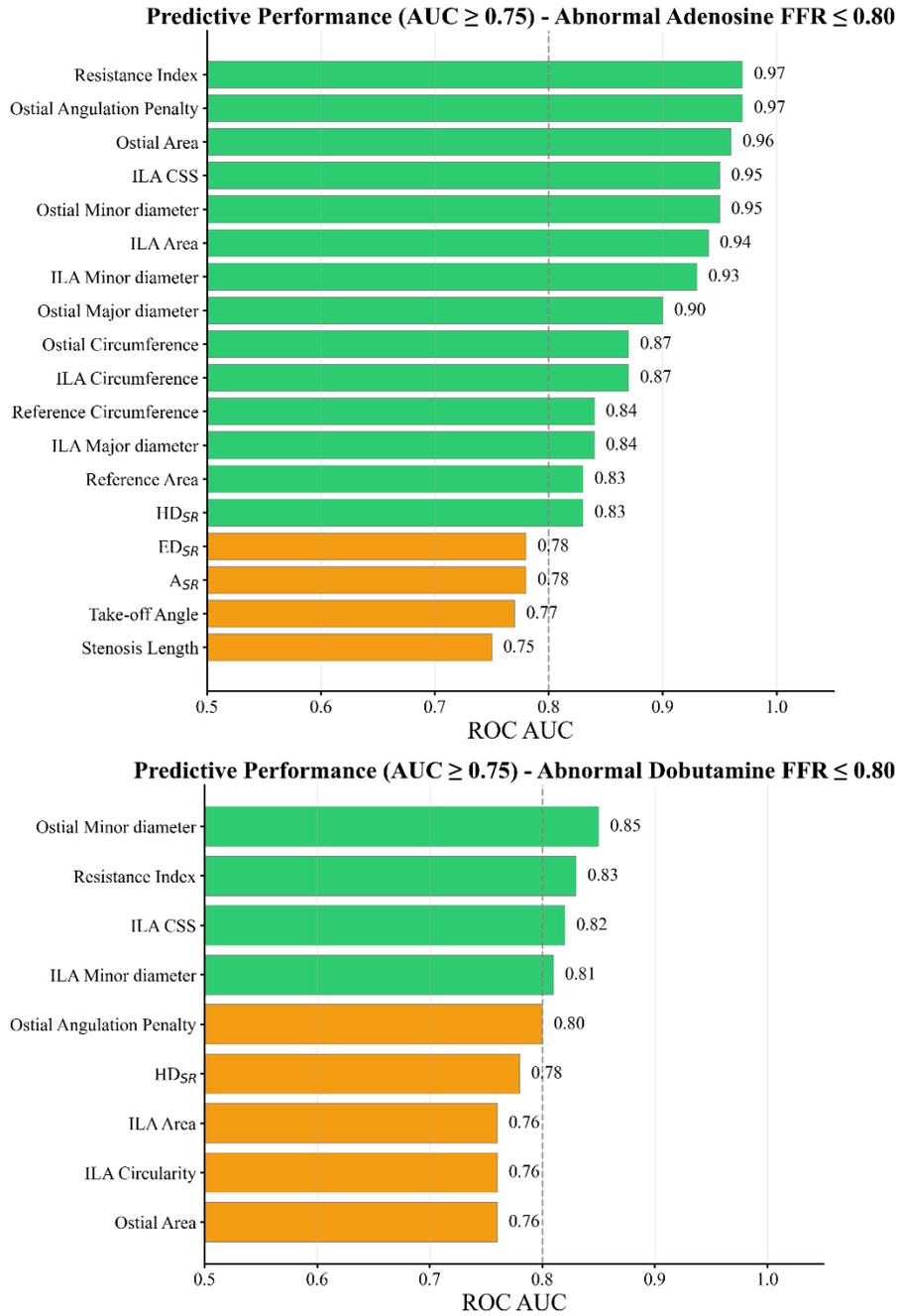

**Figure 3.** Discriminatory performance of geometric parameters for $FFR_{Adenosine}$ (top) and $FFR_{Dobutamine}$ (bottom).

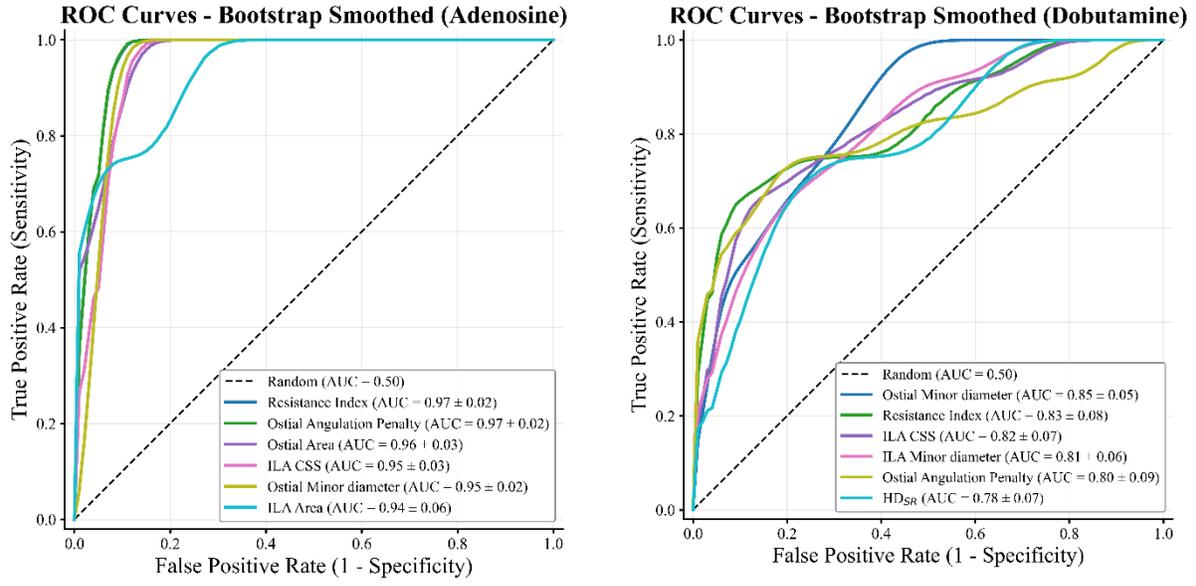

**Figure 4.** Bootstrap-smoothed ROC curves for top-performing parameters for FFRAdenosine (left) and FFRDobutamine (right).

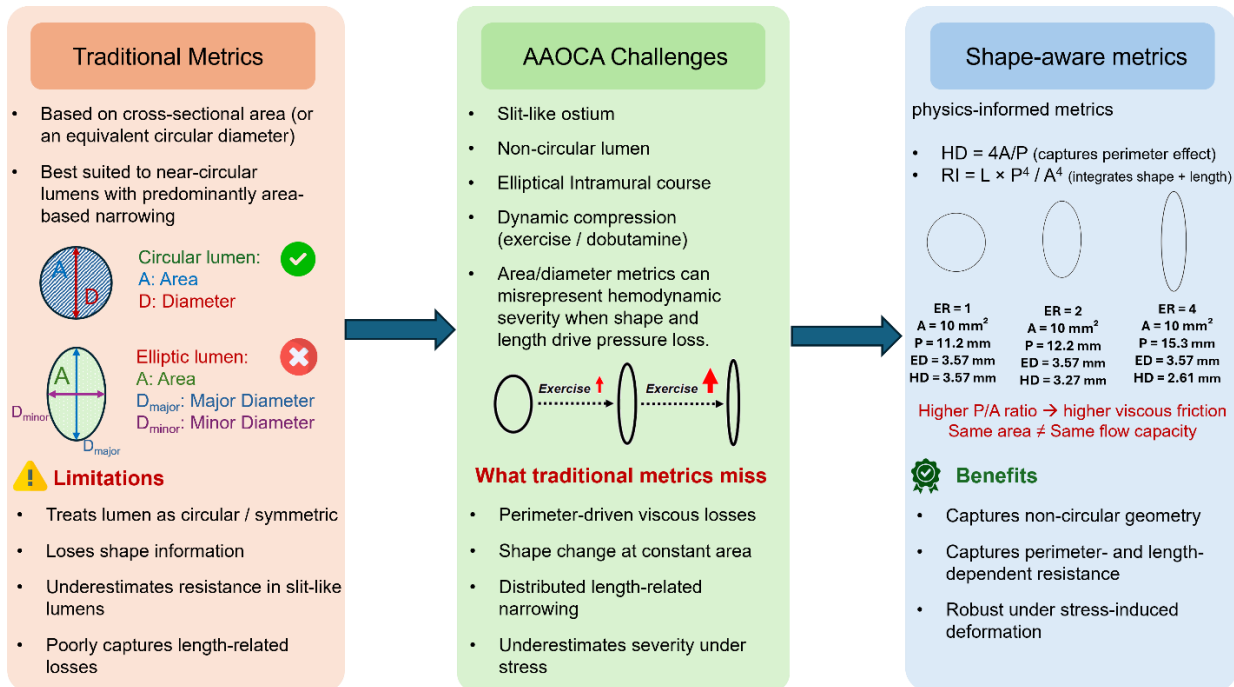

**Figure 5.** Central Illustration: Conventional vs Fluid Dynamics–Informed Parameters. Conventional geometric parameters assume circular geometry; fluid dynamics–informed parameters (hydraulic diameter, resistance index) capture perimeter-dependent resistance in elliptical R-AAOCA lumens.

# Tables

**Table 1.** Summary of CCTA-Derived Geometric Parameters

| Parameter | Symbol | Formula | Unit | Description |
|---|---|---|---|---|
| **Conventional Geometric Parameters** | | | | |
| Stenosis Length | $L$ | - | mm | Distance from ostium to intramural exit |
| Intramural Lumen Area | $A_{ILA}$ | - | mm² | Cross-sectional area at maximal stenosis |
| Minimum Lumen Diameter | $D_{minor}$ | - | mm | Shortest diameter at ILA |
| Maximum Lumen Diameter | $D_{major}$ | - | mm | Longest diameter at ILA |
| Reference Area | $A_{ref}$ | - | mm² | Normal vessel area 10 mm distal |
| Take-off Angle | $\theta$ | - | ° | Angle between coronary and aortic wall |
| Area Stenosis Ratio | $A_{SR}$ | $100\times(1-A_{ILA}/A_{ref})$ | % | Percent area reduction |
| Effective Diameter Stenosis Ratio | $ED_{SR}$ | $100\times(1-D_{eff,ILA}/D_{eff,ref})$ | % | Traditional ratio (circular assumption) |
| **Fluid Dynamics–Informed Parameters** | | | | |
| **Shape Descriptors** | | | | |
| Elliptic Ratio | $ER$ | $D_{major}/D_{minor}$ | - | Major-to-minor axis; 1.0 = circular |
| Circularity | $C$ | $4\pi A/P^2$ | - | Compactness; 1.0 = perfect circle |
| Hydraulic Diameter | $D_h$ | $4A/P$ | mm | Shape-adjusted diameter for resistance |
| Hydraulic Diameter Stenosis Ratio | $HD_{SR}$ | $100\times(1-D_{h,ILA}/D_{h,ref})$ | % | Shape-aware stenosis ratio |
| **Resistance-Based Indices** | | | | |
| Ostial Angulation Penalty | OAP | $A_{ost}/A_{ref}\times(1+\tan\theta)$ | - | Ostial lumen preservation weighted by take-off angle |
| Resistance Index | RI | $L\times P^4/A^4$ | mm⁻³ | Poiseuille-based ($\propto L/D_h^4$) |
| Comprehensive Stenosis Score | CSS | $RI\times ER\times[1+\sin\theta]$ | mm⁻³ | Integrated resistance, shape, angle |

*Abbreviations:* $A$, area; $D$, diameter; $D_{eff} = \sqrt{(4A/\pi)}$; ILA, intramural lumen area; $L$, length; $P$, perimeter.

**Table 2.** Correlation and Linear Regression Analysis of Geometric Parameters with FFR by Stress Modality

| Parameter | r (95% CI) | $R^2$ | β (SE) | p-value |
|---|---|---|---|---|
| **$FFR_{Adenosine}$** | | | | |
| Resistance Index | −0.67 (−0.77, −0.53) | 0.45 | −0.001 (0.000) | <0.001 |
| ILA CSS | −0.53 (−0.67, −0.35) | 0.27 | −0.001 (0.000) | <0.001 |
| Ostial Area | 0.43 (0.23, 0.59) | 0.18 | 0.013 (0.003) | <0.001 |
| Ostial Angulation Penalty | 0.37 (0.16, 0.54) | 0.13 | 0.076 (0.027) | <0.001 |
| Ostial Minor Diameter | 0.28 (0.07, 0.47) | 0.08 | 0.041 (0.016) | 0.011 |
| ILA Area | 0.35 (0.14, 0.53) | 0.12 | 0.012 (0.004) | 0.002 |
| **$FFR_{Dobutamine}$** | | | | |
| Ostial Minor Diameter | 0.47 (0.29, 0.63) | 0.23 | 0.091 (0.019) | <0.001 |
| Resistance Index | −0.66 (−0.77, −0.51) | 0.43 | −0.005 (0.001) | <0.001 |
| ILA CSS | −0.59 (−0.72, −0.43) | 0.35 | −0.002 (0.000) | <0.001 |
| Ostial Angulation Penalty | 0.47 (0.28, 0.62) | 0.22 | 0.127 (0.027) | <0.001 |
| ILA Minor Diameter | 0.50 (0.32, 0.65) | 0.25 | 0.107 (0.021) | <0.001 |
| Hydraulic Diameter Stenosis Ratio | −0.43 (−0.59, −0.23) | 0.18 | −0.284 (0.067) | <0.001 |

*Note:* Parameters ordered by ROC AUC within each stress modality. Abbreviations: β, regression coefficient (change in FFR per unit increase in parameter); CI, confidence interval; ILA, intramural lumen area; r, Pearson correlation coefficient; $R^2$, coefficient of determination; SE, standard error. Confidence intervals for r derived using Fisher's z-transformation.

**Table 3.** Univariable Logistic Regression for Predicting Hemodynamically Significant Stenosis (FFR ≤ 0.80) by Stress Modality

| Parameter | β | SE | z | OR (95% CI) | p-value |
|---|---|---|---|---|---|
| **FFR$_{Adenosine}$** | | | | | |
| Resistance Index | 1.86 | 0.71 | 2.62 | 6.39 (1.6 - 25.59) | 0.006 |
| ILA CSS | 0.96 | 0.35 | 2.76 | 2.61 (1.32 - 5.16) | 0.009 |
| Ostial Minor Diameter | −4.53 | 2.02 | −2.24 | 0.01 (0.0 - 0.57) | 0.025 |
| ILA Minor Diameter | −3.14 | 1.27 | −2.46 | 0.04 (0.0 - 0.53) | 0.014 |
| Ostial Angulation Penalty | −5.08 | 2.1 | −2.41 | 0.01 (0.0 - 0.38) | 0.016 |
| Ostial Area | −5.07 | 2.23 | −2.28 | 0.01 (0.0 - 0.49) | 0.023 |
| **FFR$_{Dobutamine}$** | | | | | |
| Resistance Index | 1.83 | 0.54 | 3.39 | 6.26 (2.17 - 18.06) | 0.001 |
| ILA CSS | 1.24 | 0.43 | 2.87 | 3.47 (1.48 - 8.12) | 0.002 |
| Ostial Minor Diameter | −2.15 | 0.68 | −3.18 | 0.12 (0.03 - 0.44) | 0.001 |
| ILA Minor Diameter | −1.63 | 0.55 | −2.95 | 0.20 (0.07 - 0.58) | 0.003 |
| Ostial Angulation Penalty | −1.74 | 0.66 | −2.89 | 0.18 (0.05 - 0.57) | 0.004 |
| Hydraulic Diameter Stenosis Ratio | 1.36 | 0.49 | 2.80 | 3.91 (1.51 - 10.16) | 0.005 |

*Note:* Parameters were z-score standardized; OR reflects change in odds per 1 standard deviation increase. Abbreviations: β, regression coefficient; CI, confidence interval; CSS, comprehensive stenosis score; ILA, intramural lumen area; OR, odds ratio; SE, standard error.

**Table 4.** Comparative ROC Performance Across Stress Modalities

| Parameter | AUC (FFR$_{Adenosine}$) | AUC (FFR$_{Dobutamine}$) | Δ AUC | Interpretation |
|---|---|---|---|---|
| Resistance Index | 0.97 | 0.83 | −0.14 | Excellent → Good |
| Ostial Angulation Penalty | 0.97 | 0.80 | −0.17 | Excellent → Good |
| ILA CSS | 0.95 | 0.82 | −0.13 | Excellent → Good |
| Ostial Minor Diameter | 0.95 | 0.85 | −0.10 | Excellent → Good |
| Ostial Area | 0.96 | 0.76 | −0.20 | Excellent → Fair |
| ILA Area | 0.94 | 0.76 | −0.18 | Excellent → Fair |
| ILA Minor Diameter | 0.93 | 0.81 | −0.12 | Excellent → Good |
| Hydraulic Diameter Stenosis Ratio | 0.83 | 0.78 | −0.05 | Good → Fair |

*Abbreviations:* AUC, area under the receiver operating characteristic curve; CSS, comprehensive stenosis score; ILA, intramural lumen area; Δ, difference. Interpretation based on Hosmer-Lemeshow criteria: Excellent (AUC ≥ 0.9), Good (0.8–0.9), Fair (0.7–0.8).

**Table 5.** Optimal Cutoff Values for Identifying Hemodynamically Significant Stenosis (FFR ≤ 0.80)

| Parameter | Cutoff | Sensitivity | Specificity | PPV | NPV | AUC |
|---|---|---|---|---|---|---|
| **FFR$_{Adenosine}$ ≤ 0.80 in 5/81 patients (6.2%)** | | | | | | |
| Resistance Index | ≥ 406.42 | 100% | 95% | 50% | 100% | 0.97 |
| CSS | ≥ 1068.29 | 100% | 92% | 40% | 100% | 0.95 |
| Ostial Area | ≤ 4.14 mm² | 100% | 89% | 38% | 100% | 0.96 |
| Ostial Minor Diameter | ≤ 1.10 mm | 100% | 92% | 45% | 100% | 0.95 |
| ILA Area | ≤ 4.10 mm² | 100% | 78% | 24% | 100% | 0.94 |
| **FFR$_{Dobutamine}$ ≤ 0.80 in 16/81 patients (19.8%)** | | | | | | |
| Ostial Minor Diameter | ≤ 1.60 mm | 100% | 57% | 43% | 100% | 0.85 |
| ILA Minor Diameter | ≤ 1.60 mm | 90% | 56% | 40% | 94% | 0.81 |
| Resistance Index | ≥ 284.59 | 69% | 94% | 64% | 93% | 0.83 |
| CSS | ≥ 796.84 | 67% | 91% | 57% | 94% | 0.82 |
| Ostial Angulation Penalty | ≥ 0.62 | 75% | 84% | 43% | 95% | 0.8 |

*Note:* Cutoffs determined using Youden's J statistic (maximizing sensitivity + specificity − 1). Abbreviations: AUC, area under the curve; CSS, comprehensive stenosis score; FFR, fractional flow reserve; ILA, intramural lumen area; NPV, negative predictive value; OAP, ostial angulation penalty; PPV, positive predictive value.

# Supplementary material

## Detailed Description of Geometric Parameters

**Conventional Geometric Parameters**

**Stenosis Length.** Stenosis length ($L$, mm) was defined as the longitudinal extent of the abnormal proximal segment, measured as the distance from the ostium to the distal termination of the intramural segment, identified on CCTA as the point where the anomalous vessel exits the aortic wall and assumes a normal extramural course.

**Intramural Lumen Area.** Intramural lumen area ($A_{ILA}$, mm²) was defined as the cross-sectional lumen area at the point of maximal stenosis within the intramural segment, representing the most severe geometric constriction.

**Minimum and Maximum Lumen Diameters.** The minimum lumen diameter ($D_{minor}$, mm) represents the shortest diameter of the lumen at ILA and the most restrictive dimension for flow. The maximum lumen diameter ($D_{major}$, mm) defines the longest axis of the elliptical cross-section.

**Reference Segment.** A 5-mm segment of normal-appearing vessel ($A_{ref}$, mm²) located 10 mm distal to the intramural segment termination, where the vessel assumed normal circular morphology, was used as the reference for all stenosis ratio calculations.

**Take-off Angle.** The acute angle ($\theta$, degrees) between the anomalous coronary artery and the aortic wall at its origin, measured in the plane of the aortic root. Acute angles contribute to flow separation and entrance losses.

**Area Stenosis Ratio.** The relative area reduction compared to the reference segment: $A_{SR} = 100 \times (1 - A_{ILA} / A_{ref})$, expressed as percentage.

**Effective Diameter Stenosis Ratio.** A conventional diameter-based metric assuming circular geometry: $ED_{SR} = 100 \times (1 - D_{eff,ILA} / D_{eff,ref})$, where $D_{eff} = \sqrt{4A/\pi}$.

**Fluid Dynamics–Informed Parameters**

**Shape Descriptors**

**Elliptic Ratio.** A dimensionless shape descriptor quantifying luminal eccentricity: $ER = D_{major} / D_{minor}$. Values approaching 1.0 indicate circular geometry; higher values indicate progressive elliptical distortion characteristic of compressed intramural segments. ER was calculated at both the ostium ($ER_{ostial}$) and ILA ($ER_{ILA}$) to capture site-specific geometric deformation.

**Circularity.** A dimensionless compactness metric quantifying deviation from circular geometry: $C = 4\pi A / P^2$. For a perfect circle, $C = 1.0$; lower values indicate increasing perimeter relative to area, reflecting shape irregularity or elliptical distortion. Unlike ER, circularity penalizes any perimeter inflation including irregular or jagged luminal boundaries. Circularity was calculated at both the ostium and ILA.

**Hydraulic Diameter.** A shape-adjusted diameter derived from fluid dynamics principles: $D_h = 4A / P$. For a perfect circle, $D_h$ equals the geometric diameter. For elliptical or irregular cross-sections, $D_h$ is smaller than the area-equivalent diameter because the larger perimeter increases wall shear and frictional resistance. This parameter captures the disproportionate flow limitation imposed by slit-like geometries.

**Hydraulic Diameter Stenosis Ratio.** A shape-aware stenosis metric: $HD_{SR} = 100 \times (1 - D_{h,ILA} / D_{h,ref})$. This accounts for both area reduction and perimeter-dependent resistance changes in non-circular conduits.

**Resistance-Based Indices**

**Ostial Angulation Penalty.** A composite index incorporating take-off angle effects: $OAP = (A_{ostial} / A_{ref}) \times (1 + \tan(\theta))$. Higher OAP values indicate a larger ostial cross-sectional area relative to the reference segment (greater lumen preservation at the coronary origin), amplified by more perpendicular take-off angles. The $\tan(\theta)$ term scales with angle, reaching its maximum contribution as $\theta$ approaches 90°. For interpretability, higher OAP values indicate lower hemodynamic risk (i.e., greater ostial lumen preservation relative to the reference segment and therefore higher expected FFR).

**Resistance Index.** A fluid dynamic–derived parameter incorporating stenosis length and perimeter-to-area relationship: $RI = L \times P^4 / A^4$. This formulation is grounded in Poiseuille's law for laminar flow, where resistance scales with length and inversely with the fourth power of hydraulic diameter. Since $D_h = 4A/P$, the $P^4/A^4$ term ensures proper scaling with resistance $\propto L/D_h^4$. Values were calculated at the site of maximal stenosis (ILA).

**Comprehensive Stenosis Score.** An integrated index combining resistance, shape, and angulation effects: $CSS = RI \times ER_{ILA} \times [1 + \sin(\theta)]$. This formulation synthesizes three pathophysiologic mechanisms: (1) geometric resistance through RI, (2) shape-dependent viscous losses through the ER penalty, and (3) entrance effects from acute take-off angles.

All geometric parameters were computed from CCTA-derived measurements of area, perimeter, and lumen axes at standardized locations (ostium, ILA, and reference segment). Parameters requiring ratios used the reference segment as denominator to normalize for vessel size.

## A short review of the parameters in the cohort

Supplementary Table S1 presents the complete descriptive statistics for all geometric parameters measured in the study cohort. The table includes mean±SD, median, minimum, and maximum values for demographic characteristics (age, body mass index) and all CCTA-derived geometric parameters. Parameters are organized into the two categories defined in the main manuscript: conventional geometric parameters and fluid dynamics–informed parameters. The stenosis ratios (HDSR, EDSR, ASR) represent the relationship between stenotic and reference segment dimensions using hydraulic diameter, effective diameter, and area approaches, respectively. The fluid dynamics–informed parameters including the RI, OAP, and CSS integrate multiple anatomical features to quantify cumulative hemodynamic burden. The wide range of values observed for most parameters reflects the heterogeneous anatomical presentation of R-AAOCA in the study population.

**Supplementary Table S1.** Baseline Clinical and Geometric Characteristics of the Study Population

| Parameter | Mean±SD | Median | Range |
|---|---|---|---|
| **Clinical Characteristics** | | | |
| Age (years) | 52.3±12.0 | 50.5 | 22 - 81 |
| Body mass index (kg/m²) | 27.0±5.1 | 26.0 | 18.6 - 38.8 |
| **Conventional Geometric Parameters** | | | |
| Ostial area (mm²) | 6.69±2.27 | 6.59 | 1.34 - 13.0 |
| Ostial circumference (mm) | 10.92±1.95 | 10.95 | 4.9 - 16.7 |
| Ostial major diameter (mm) | 4.58±0.92 | 4.60 | 1.4 - 7.1 |
| Ostial minor diameter (mm) | 1.69±0.47 | 1.60 | 0.9 - 3.1 |
| ILA area (mm²) | 5.55±1.94 | 5.00 | 1.27 - 10.4 |
| ILA circumference (mm) | 9.64±1.76 | 9.60 | 4.9 - 14.1 |
| ILA major diameter (mm) | 3.89±0.81 | 3.90 | 1.4 - 5.8 |
| ILA minor diameter (mm) | 1.66±0.42 | 1.60 | 0.9 - 2.8 |
| Reference area (mm²) | 11.62±3.35 | 11.00 | 5.0 - 22.0 |
| Reference circumference (mm) | 11.84±1.75 | 11.75 | 8.0 - 17.0 |
| Take-off angle (°) | 22.9±11.6 | 24.8 | 0 - 43 |
| Stenosis length (mm) | 10.41±4.20 | 10.30 | 0 - 21.5 |
| Area stenosis ratio ($A_{SR}$) | 50±17 | 54 | 0 - 87 |
| Effective diameter stenosis ratio ($ED_{SR}$) | 31±12 | 32 | 0 - 64 |
| **Fluid Dynamics–Informed Parameters** | | | |
| Resistance index (mm$^{-3}$) | 179.91±261.02 | 116.59 | 0 - 1994.4 |
| ILA CSS (mm$^{-3}$) | 497.05±693.83 | 306.75 | 0 - 4186.34 |
| Ostial angulation penalty | 0.87±0.34 | 0.82 | 0.21 - 2.32 |
| Ostial ellipticity | 2.89±0.85 | 2.80 | 1.0 - 5.3 |
| Ostial circularity | 0.70±0.14 | 0.69 | 0.40 - 1.0 |
| ILA ellipticity | 2.46±0.72 | 2.40 | 1.0 - 4.8 |
| ILA circularity | 0.74±0.13 | 0.72 | 0.44 - 1.0 |
| Hydraulic diameter stenosis ratio ($HD_{SR}$) | 40±14 | 44 | 0 - 71 |

*Note:* Values are mean ± standard deviation unless otherwise specified. Abbreviations: $A_{SR}$, area stenosis ratio; CSS, comprehensive stenosis score; $ED_{SR}$, effective diameter stenosis ratio; $HD_{SR}$, hydraulic diameter stenosis ratio; ILA, intramural lumen area.

## Standard ROC Curves for Top-Performing Parameters

Figure S1 presents standard ROC curves for the six highest-performing geometric parameters for $FFR_{Adenosine}$ and $FFR_{Dobutamine}$, respectively. Each curve illustrates the trade-off between sensitivity (true positive rate) and 1-specificity (false positive rate) across all possible classification thresholds for identifying hemodynamically significant stenosis (FFR ≤ 0.80). The diagonal reference line represents random classification (AUC = 0.50). For $FFR_{Adenosine}$, the RI achieved near-perfect discrimination with curve approaching the upper-left corner, reflecting its exceptional ability to distinguish patients with and without hemodynamic significance. For $FFR_{Dobutamine}$, the curves demonstrate attenuated but clinically meaningful discrimination, with ostial minor diameter showing the most favorable sensitivity-specificity profile. The separation between curves illustrates the differential predictive performance of fluid dynamics–informed versus conventional geometric parameters across stress modalities.

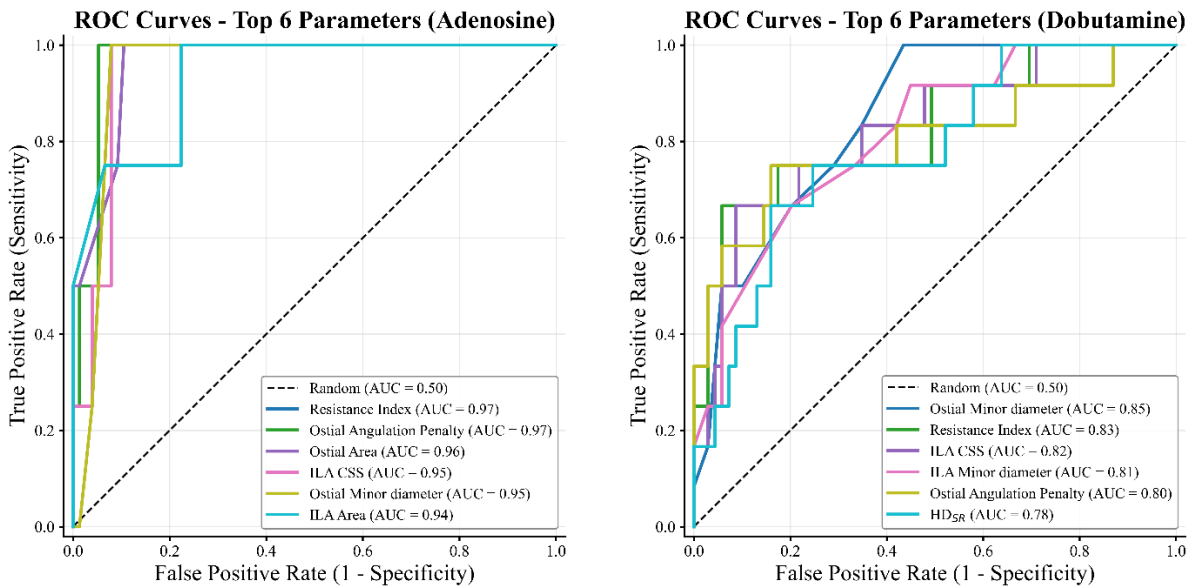

**Figure S1.** Standard ROC curves for top-performing parameters for $FFR_{Adenosine}$ (left) and $FFR_{Dobutamine}$ (right).

Abbreviations: AUC, area under the ROC curve; CSS, comprehensive stenosis score; $FFR_{Adenosine}$, fractional flow reserve during adenosine stress; $FFR_{Dobutamine}$, fractional flow reserve during dobutamine stress; ILA, intramural lumen area; OAP, ostial angulation penalty; RI, resistance index; ROC, receiver operating characteristic.

**Inter-Parameter Correlation Matrices**

Supplementary Figures S2 and S3 display triangular correlation heatmaps illustrating Pearson correlation coefficients among all geometric parameters and $FFR_{Adenosine}$ and $FFR_{Dobutamine}$ values, respectively. Color intensity indicates correlation strength (red = positive, blue = negative), with values ranging from −1.0 to +1.0. Several patterns merit attention: (1) Strong positive correlations exist among conventional geometric parameters (areas, diameters, circumferences) reflecting their shared dependence on vessel size; (2) fluid dynamics–informed parameters (RI and CSS) show strong negative correlations with FFR, confirming their inverse relationship with coronary flow; (3) ER and circularity metrics demonstrate expected inverse relationships (higher ellipticity corresponds to lower circularity); (4) The correlation structure is largely preserved across stress modalities, supporting the stability of geometric relationships despite different hemodynamic conditions. These matrices also reveal collinearity among related parameters, informing interpretation of univariable analyses and highlighting which parameters provide independent versus redundant information.

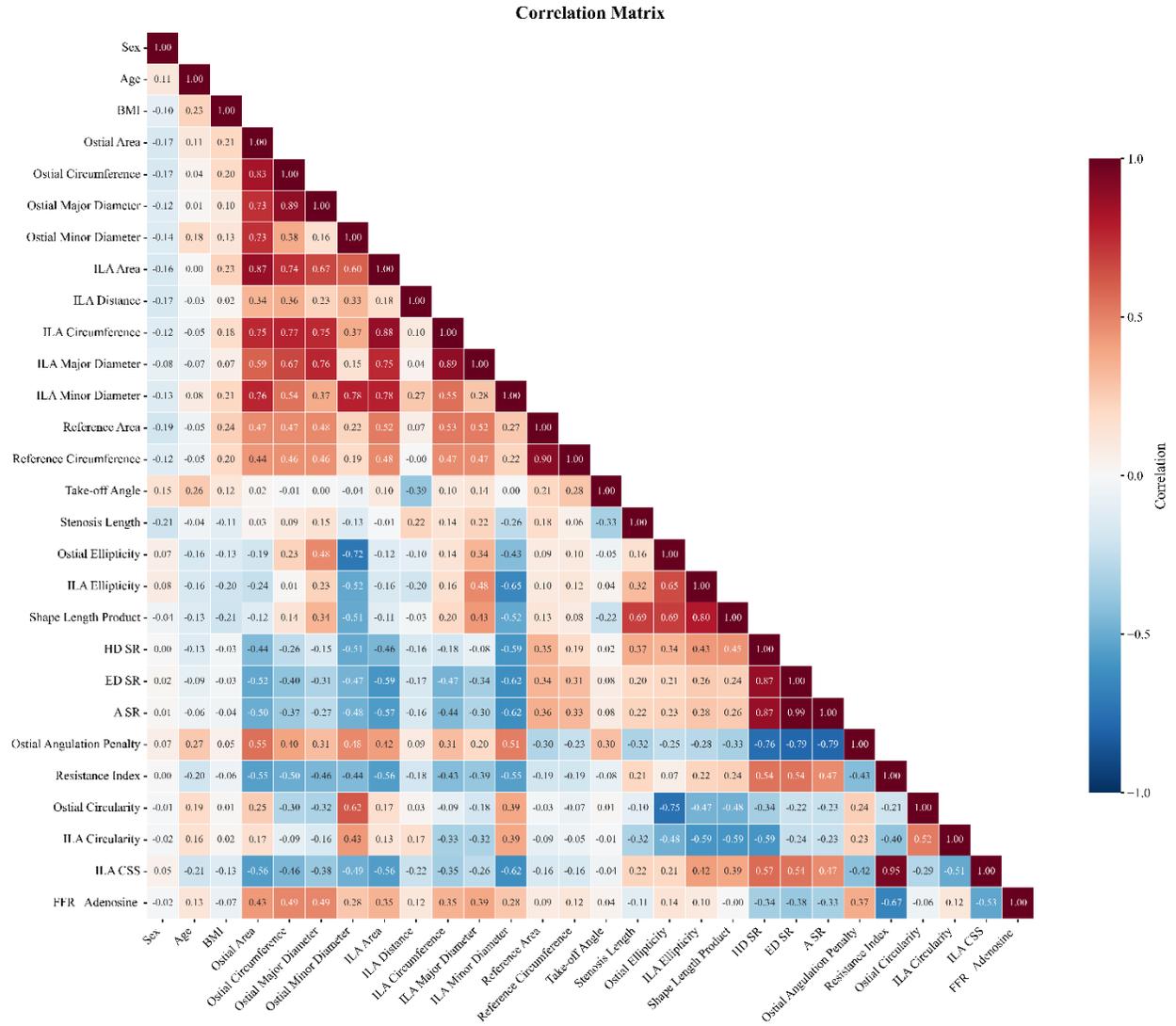

**Figure S2:** Correlation heatmap for FFR$_{Adenosine}$

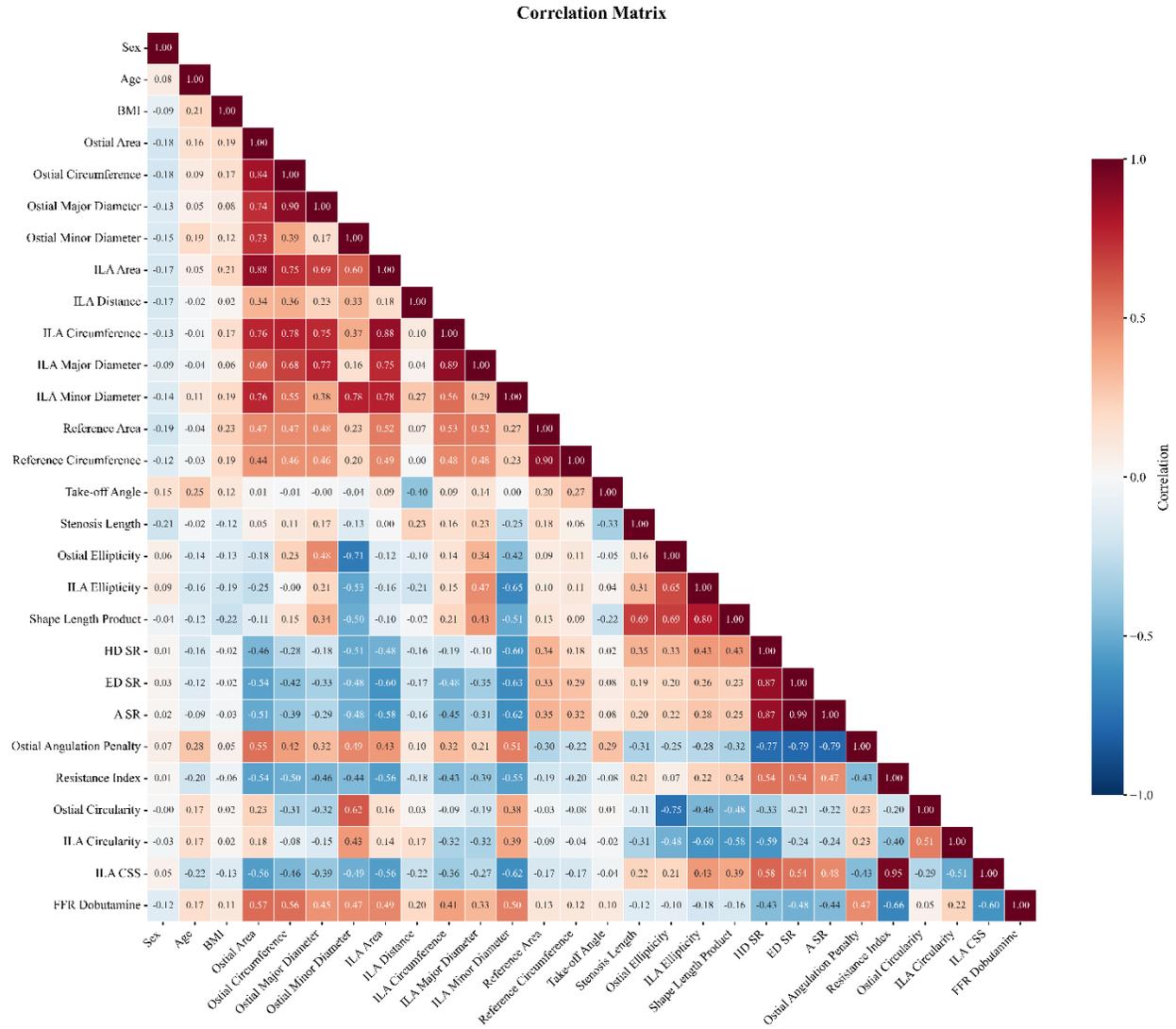

**Figure S3:** Correlation heatmap **for FFR$_{Dobutamine}$**

## Statistical Comparison of ROC Curves

Pairwise comparisons of discriminatory performance between geometric parameters were performed using DeLong's test for comparing correlated ROC curves. This non-parametric method, originally described by DeLong et al. (*Biometrics* 1988;44:837-845), accounts for the correlation between area under the curve (AUC) estimates derived from the same sample by computing a variance-covariance matrix using the theory of generalized U-statistics. The test statistic follows an asymptotic normal distribution under the null hypothesis of equal AUCs, enabling valid hypothesis testing for differences in diagnostic accuracy between correlated predictors.

The DeLong test was applied to all pairwise combinations of parameters achieving AUC ≥ 0.75 under each stress condition ($FFR_{Adenosine}$ and $FFR_{Dobutamine}$). For transparency, the complete set of pairwise test results was reviewed in the analysis, while Supplementary Tables S2 and S3 present a curated subset of the most clinically relevant/top-performing comparisons to improve readability. For each comparison, we report the AUC values for both parameters, the difference in AUC (Δ AUC), the DeLong z-statistic, and the two-sided p-value. Statistical significance was defined as p < 0.05. Given the exploratory nature of these comparisons, no adjustment for multiple testing was applied; results should be interpreted in the context of effect sizes and confidence intervals rather than strict hypothesis testing.

**Supplementary Results: Summary of AUC Significance Testing**

$FFR_{Adenosine}$: A total of 28 pairwise comparisons were performed among parameters with AUC ≥ 0.75 for $FFR_{Adenosine}$. No statistically significant differences were observed (all p > 0.05). The smallest p-value was 0.104 (RI [AUC = 0.97] vs ILA CSS [AUC = 0.95], Δ AUC = 0.02, z = 1.62). This finding indicates that the top-performing parameters, including RI, CSS, ostial area, ostial minor diameter, ILA area, and OAP, provide statistically comparable discriminatory performance for detecting hemodynamically significant stenosis by $FFR_{Adenosine}$. Supplementary Table S2 shows a subset of representative comparisons among the highest-performing parameters (primarily AUC ≥ 0.90), whereas the statement regarding 28 comparisons refers to the full set of tested pairwise comparisons with AUC ≥ 0.75

$FFR_{Dobutamine}$: A total of 28 pairwise comparisons were performed among parameters with AUC ≥ 0.75 for $FFR_{Dobutamine}$. One statistically significant difference was observed: ostial minor diameter demonstrated significantly higher AUC than ostial area (0.85 vs 0.76, Δ AUC = 0.09, z = 2.04, p = 0.041). The next smallest p-value was 0.052 (RI [AUC = 0.83] vs ILA Area [AUC = 0.76], Δ AUC = 0.07, z = 1.95). All other pairwise comparisons were non-significant, suggesting that among the top-performing parameters (excluding the ostial minor diameter vs ostial area comparison), discriminatory performance is broadly equivalent for $FFR_{Dobutamine}$. Supplementary Table S3 presents a selected subset of pairwise comparisons highlighting the top-performing parameters and the key significant contrast (ostial minor diameter vs ostial area), while the reported total of 28 comparisons refers to the full set tested among parameters with AUC ≥ 0.75.

**Supplementary Table S2.** Pairwise Comparisons of Area Under the Receiver Operating Characteristic Curve Using DeLong's Test for $FFR_{Adenosine}$

| Parameter 1 | AUC | Parameter 2 | AUC | Δ AUC | z-statistic | p-value | Significance |
| --- | --- | --- | --- | --- | --- | --- | --- |
| Resistance Index | 0.97 | ILA CSS | 0.95 | 0.02 | 1.62 | 0.104 | ns |
| Resistance Index | 0.97 | Ostial Area | 0.96 | 0.01 | 0.32 | 0.752 | ns |
| Resistance Index | 0.97 | Ostial Minor Diameter | 0.95 | 0.02 | 0.81 | 0.420 | ns |
| Resistance Index | 0.97 | ILA Area | 0.94 | 0.03 | 0.69 | 0.492 | ns |
| Resistance Index | 0.97 | OAP | 0.97 | 0.00 | 0.00 | 1.000 | ns |
| Resistance Index | 0.97 | ILA Minor Diameter | 0.93 | 0.04 | 1.29 | 0.196 | ns |
| Resistance Index | 0.97 | Ostial Max Diameter | 0.9 | 0.07 | 0.94 | 0.349 | ns |
| ILA CSS | 0.95 | Ostial Area | 0.96 | 0.01 | 0.42 | 0.674 | ns |
| ILA CSS | 0.95 | Ostial Minor Diameter | 0.95 | 0.00 | 0.05 | 0.959 | ns |
| ILA CSS | 0.95 | ILA Area | 0.94 | 0.01 | 0.30 | 0.767 | ns |
| Ostial Area | 0.96 | Ostial Minor Diameter | 0.95 | 0.01 | 0.36 | 0.719 | ns |
| Ostial Area | 0.96 | ILA Area | 0.94 | 0.02 | 0.75 | 0.453 | ns |
| Ostial Area | 0.96 | OAP | 0.97 | 0.01 | 0.43 | 0.670 | ns |
| Ostial Minor Diameter | 0.95 | ILA Area | 0.94 | 0.01 | 0.20 | 0.841 | ns |
| Ostial Minor Diameter | 0.95 | OAP | 0.97 | 0.02 | 0.85 | 0.397 | ns |
| ILA Area | 0.94 | OAP | 0.97 | 0.03 | 0.70 | 0.482 | ns |

*Note:* DeLong's test was performed for all pairwise combinations among parameters with AUC ≥ 0.75 for $FFR_{Adenosine}$ (28 total comparisons). To improve readability, only a representative subset of comparisons among the highest-performing parameters (predominantly AUC ≥ 0.90) is shown here. All the comparisons displayed were non-significant. Abbreviations: AUC, area under the receiver operating characteristic curve; CSS, comprehensive stenosis score; ILA, intramural lumen area; ns, not significant; OAP, ostial angulation penalty; Δ, difference. Significance levels: * $p < 0.05$, ** $p < 0.01$, *** $p < 0.001$.

**Supplementary Table S3.** Pairwise Comparisons of Area Under the Receiver Operating Characteristic Curve Using DeLong's Test for $FFR_{Dobutamine}$

| Parameter 1 | AUC | Parameter 2 | AUC | Δ AUC | z-statistic | p-value | Significance |
|---|---|---|---|---|---|---|---|
| Ostial Minor Diameter | 0.85 | Ostial Area | 0.76 | 0.09 | 2.04 | 0.041 | * |
| Ostial Minor Diameter | 0.85 | Resistance Index | 0.83 | 0.02 | 0.27 | 0.787 | ns |
| Ostial Minor Diameter | 0.85 | ILA CSS | 0.82 | 0.03 | 0.38 | 0.701 | ns |
| Ostial Minor Diameter | 0.85 | OAP | 0.8 | 0.05 | 0.58 | 0.562 | ns |
| Ostial Minor Diameter | 0.85 | ILA Minor Diameter | 0.81 | 0.04 | 0.64 | 0.523 | ns |
| Ostial Minor Diameter | 0.85 | $HD_{SR}$ | 0.78 | 0.07 | 0.99 | 0.320 | ns |
| Ostial Minor Diameter | 0.85 | ILA Area | 0.76 | 0.09 | 1.40 | 0.160 | ns |
| Resistance Index | 0.83 | ILA Area | 0.76 | 0.07 | 1.95 | 0.052 | ns |
| ILA CSS | 0.82 | ILA Area | 0.76 | 0.06 | 1.83 | 0.068 | ns |
| Resistance Index | 0.83 | ILA CSS | 0.82 | 0.01 | 0.29 | 0.769 | ns |
| Resistance Index | 0.83 | OAP | 0.8 | 0.03 | 0.23 | 0.822 | ns |
| Resistance Index | 0.83 | ILA Minor Diameter | 0.81 | 0.02 | 0.32 | 0.753 | ns |
| ILA CSS | 0.82 | OAP | 0.8 | 0.02 | 0.19 | 0.846 | ns |
| ILA CSS | 0.82 | ILA Minor Diameter | 0.81 | 0.01 | 0.27 | 0.784 | ns |

*Note:* DeLong's test was performed for all pairwise combinations among parameters with AUC ≥ 0.75 for $FFR_{Dobutamine}$ (28 total comparisons). To improve readability, this table shows a selected subset of comparisons emphasizing top-performing parameters and the key significant contrast. Ostial minor diameter demonstrated significantly higher AUC than ostial area (p = 0.041), the only statistically significant pairwise difference observed in the full tested set. Abbreviations: $A_{SR}$, area stenosis ratio; AUC, area under the receiver operating characteristic curve; CSS, comprehensive stenosis score; $HD_{SR}$, hydraulic diameter stenosis ratio; ILA, intramural lumen area; ns, not significant; OAP, ostial angulation penalty; Δ, difference. Significance levels: * p < 0.05.

**Interpretation**

The absence of significant differences among top-performing parameters for FFRAdenosine indicates that diverse geometric parameters, including fluid dynamics–informed parameters and conventional geometric parameters, capture fundamentally similar anatomical information relevant to fixed hemodynamic obstruction. This equivalence has practical implications: clinicians may select parameters based on measurement availability and local expertise without sacrificing diagnostic accuracy. The significantly superior performance of ostial minor diameter compared with ostial area for $FFR_{Dobutamine}$ (p = 0.041) has mechanistic plausibility. Dynamic compression during exercise-mimicking stress produces slit-like luminal geometries that reduce the minor diameter disproportionately to the cross-sectional area. The minor diameter represents the most restrictive dimension for flow and therefore better captures the hemodynamic impact of dynamic compression than area-based measurements that average across the entire cross-section. This finding supports prioritizing minor diameter assessment when evaluating patients for dynamic compression risk during physiologic stress.

## Misclassification Analysis and Phenotypic Characterization

To better understand the limitations of geometric parameters in predicting hemodynamic significance, we systematically analyzed patients who were misclassified by the Resistance Index using optimal cutoff values derived from Youden's J statistic. For $FFR_{Adenosine}$ (cutoff ≥406.42 $mm^{-3}$), the Resistance Index achieved 100% sensitivity with no false negative case. For $FFR_{Dobutamine}$ (cutoff ≥284.59 $mm^{-3}$), sensitivity decreased to 69%, with 11 of 16 hemodynamically significant cases.

**False Negative Cases: Anatomically Favorable but Hemodynamically Significant**

False negative cases represent the most clinically consequential misclassifications, patients with hemodynamically significant stenosis (FFR ≤ 0.80) who would be missed by geometric screening alone. Analysis of these cases revealed a consistent anatomical phenotype characterized by relatively preserved luminal dimensions despite functional impairment (Table S4).

Compared to correctly identified (true positive) cases, false negative patients exhibited significantly larger intramural lumen areas (5.08 ± 0.89 vs 3.16 ± 1.08 mm², p = 0.003) and larger minor diameters at the ILA (1.51 ± 0.26 vs 1.14 ± 0.21 mm, p = 0.021). These patients also demonstrated shorter stenosis lengths (10.01 ± 1.86 vs 12.80 ± 2.67 mm, p = 0.080), though this difference did not reach statistical significance.

**Supplementary Table S4.** Comparison of True Positive and False Negative Cases for FFR$_{Dobutamine}$

| Parameter | True Positive (n=11) | False Negative (n=5) | p-value |
|---|---|---|---|
| Ostial Area (mm²) | 4.11±2.21 | 5.55±1.03 | 0.064 |
| ILA Area (mm²) | 3.16±1.08 | 5.08±0.89 | **0.003** |
| ILA Minor Diameter (mm) | 1.14±0.21 | 1.51±0.26 | **0.021** |
| Stenosis Length (mm) | 12.80±2.67 | 10.01±1.86 | 0.080 |
| Take-off Angle (°) | 15.64±11.45 | 22.49±10.44 | 0.166 |

*Values presented as mean ± standard deviation. P-values from Mann-Whitney U test.*

The most clinically significant false negative demonstrated FFR$_{Dobutamine}$ of 0.61 under dobutamine stress despite a Resistance Index of only 105.1 mm$^{-3}$, well below the diagnostic cutoff. This patient exhibited an ILA area of 5.22 mm² and minor diameter of 1.30 mm, dimensions that would typically be considered low-risk. Similarly, another false negative case (FFR$_{Dobutamine}$ = 0.64) had an ILA area of 5.50 mm² with a short stenosis length of only 9.0 mm. These cases highlight a phenotype where relatively preserved static anatomy does not preclude significant dynamic compression under stress conditions.

Patient NARCO_316 exemplified the extreme end of this spectrum: despite the largest ILA area in the cohort (7.00 mm²) and a minor diameter of 1.90 mm, FFR$_{Dobutamine}$ measured 0.76. This case, with a RI of only 57.3 mm$^{-3}$, would be confidently classified as low-risk by any static geometric parameter, yet demonstrated hemodynamic significance during pharmacological stress. The mechanism likely involves dynamic compression that cannot be captured by resting CCTA geometry alone.

**False Positive Cases: Anatomically Severe but Hemodynamically Compensated**

For FFR$_{Dobutamine}$, only four patients were classified as false positives, demonstrating high RI values (≥284.59 mm$^{-3}$) despite normal FFR. These cases exhibited anatomical features that would typically predict hemodynamic compromise: small ILA areas (3.00–4.65 mm²), small minor diameters (1.00–1.40 mm), and in some cases, long stenosis lengths (up to 20.0 mm).

One particularly notable case demonstrated marked discordance: despite a RI of 549.6 mm$^{-3}$ and ILA minor diameter of only 1.00 mm, FFR$_{Dobutamine}$ measured 0.91. Similarly, for another case (RI = 605.5 mm$^{-3}$, ILA = 3.00 mm², stenosis length = 15.5 mm) maintained FFR$_{Dobutamine}$ of 0.82. These cases may reflect physiological compensation not captured by resting geometry, such as preserved coronary flow reserve or other patient-specific hemodynamic factors; however, the present study was not designed to determine the mechanism of discordance.

**Clinical Implications of Misclassification Patterns**

The asymmetric distribution of misclassification, more false negatives for FFRDobutamine than FFRAdenosine, has important clinical implications. FFRAdenosine primarily reflects microvascular resistance reduction, creating a relatively fixed demand that correlates well with static geometry. In contrast, FFRDobutamine reflects dynamic conditions (increased heart rate, contractility, and cardiac output) that may produce compression patterns not reflected in resting anatomy. The higher false negative rate for FFRDobutamine (31% vs 0% for FFRAdenosine) suggests that patients with "anatomically favorable" R-AAOCA may still experience clinically relevant ischemia during physiological stress states that better simulate exercise.

These findings support a clinical approach where geometric parameters serve as initial screening tools rather than definitive arbiters. Patients below diagnostic thresholds, particularly those with preserved luminal dimensions but clinical symptoms or concerning stress test findings, may warrant invasive assessment. The 100% sensitivity of the RI for $FFR_{Adenosine}$ suggests its utility for ruling out hemodynamic significance in low-risk patients, while the 69% sensitivity for $FFR_{Dobutamine}$ underscores the need for functional testing in equivocal cases.

In summary, misclassification analysis reveals two distinct phenotypes: (1) patients with preserved static geometry who experience dynamic compression under stress (false negatives), and (2) patients with severe anatomical narrowing who maintain adequate perfusion through compensatory mechanisms (false positives). Recognition of these phenotypes informs the appropriate integration of geometric parameters within a comprehensive R-AAOCA evaluation framework.